\documentstyle[11pt,fleqn]{article} 
\begin{document}
\baselineskip=8mm

\noindent
{\Large \bf Gamma-ray bursts: postburst evolution of fireballs}

\vspace*{20mm}
\noindent
{\bf Y.F. Huang$^{1}$, Z.G. Dai$^{1}$, D.M. Wei$^{2}$ and T. Lu$^{3,1,4}$}

\noindent
{\sl $^{1}$ Department of Astronomy, Nanjing University, 
Nanjing 210093, China}

\noindent
{\sl $^{2}$ Purple Mountain Observatory, Chinese Academy of Sciences,
Nanjing 210008, China }

\noindent
{\sl $^{3}$ CCAST(World Laboratory) P.O.Box.8730, Beijing 100080, China}

\noindent
{\sl $^{4}$ LCRHEA, IHEP, CAS, Beijing, China}

\vspace{10.0mm} 

\noindent
Running head: Y.F. Huang et al.: Postburst evolution of fireballs 

\vspace{1cm}

\noindent
{\em Send offprint requests to:} \\ \mbox{} \hspace {0.5cm} Z.G. Dai$^{\dag}$, 
Dept. of Astronomy,
Nanjing University, Nanjing, 210093, China 

\noindent
$^{\dag}$Tel. No.: 86 25 3215880 

\noindent
$^{\dag}$Telefax: 86 25 3326467

\noindent
$^{\dag}$E-mail: daizigao@public1.ptt.js.cn

%\noindent
%Thesaurus code numbers: 02.04.1, 02.05.1, 08.09.3 

\vspace{50mm}

\noindent
All correspondence please send to: Z.G. Dai 

\newpage

\noindent
{\Large \bf Gamma-ray bursts: postburst evolution of 
fireballs}

\vspace*{20mm}
\noindent
{\bf Y.F. Huang$^{1}$, Z.G. Dai$^{1}$, D.M. Wei$^{2}$ and T. Lu$^{3,1,4}$}

\noindent
{\sl $^{1}$ Department of Astronomy, Nanjing University, 
Nanjing 210093,  China}

\noindent
{\sl $^{2}$ Purple Mountain Observatory, Chinese Academy of Sciences,
Nanjing 210008 }

\noindent
{\sl $^{3}$ CCAST(World Laboratory) P.O.Box.8730, Beijing 100080, China}

\noindent
{\sl $^{4}$ LCRHEA, IHEP, CAS, Beijing, China}
\vspace{20mm}

\noindent
{\bf  ABSTRACT  \\ \\}
The postburst evolution of fireballs that produce $\gamma$-ray
bursts is studied, assuming the expansion of fireballs to be adiabatic
and relativistic. Numerical results as well as an approximate
analytic solution for the evolution are presented.
Due to adoption of a new relation among
$t$, $R$ and $\gamma$ (see the text), our results differ markedly from 
the previous studies. Synchrotron radiation from the shocked 
interstellar medium is attentively calculated, using a convenient
set of equations. The observed X-ray flux of GRB afterglows
can be reproduced easily. Although the optical afterglows seem much
more complicated, our results can still present a rather satisfactory
approach to observations. It is also found that the expansion will no 
longer be highly relativistic about 4 days after the main GRB.
We thus suggest that the marginally relativistic
%subrelativistic 
phase of the expansion should be investigated so as to check the 
afterglows observed a week or more later. 

\vspace{10mm}

\newpage

\section{\bf INTRODUCTION}
Since their discovery nearly thirty years ago (Klebesadel, Strong \& Olson
1973), $\gamma$-ray bursts (GRBs) have made one of the biggest mysteries
in astrophysics (Fishman \& Meegan 1995), primarily because they have
remained invisible at longer wavelengths.
The situation started to change drastically in 1997 due to the
Italian-Dutch BeppoSAX satellite (Piro et al. 1995), whose prominent
observations led to the discoveries of multi-wavelength counterparts
of several GRBs: GRB970111 (Costa et al. 1997a), 
GRB970228 (Costa et al. 1997b), GRB970402 (Feroci et al. 1997;
Heise et al. 1997) and GRB970508 (Costa et al. 1997d).
The corresponding afterglows in X-rays
(GRB970228: Costa et al. 1997c,e; GRB970402: Piro et al. 1997a; GRB970508:
Piro et al. 1997b), in optical band (GRB970228: Groot et al. 1997,
van Paradijs et al. 1997, Sahu et al. 1997, and Galama et al. 1997;
GRB970508: Bond 1997), and in radio band (GRB970508: Frail et al. 1997)
were observed with unprecedented enthusiasm and collaboration, and the
results are exciting. GRB970228 seems to be associated with a faint
galaxy (van Paradijs et al. 1997), and the redshift of the optical
counterpart of GRB970508 was even determined to be between 0.835
and 2 (Metzger et al. 1997).
Very recently it is also reported that X-ray afterglows of
GRB970616 (Marshall et al. 1997a; Murakami et al. 1997a),
GRB970828 (Remillard et al. 1997) and possibly GRB970815 (IAU Circular
No. 6718) were observed due to efficacious cooperations between
Compton Gamma-Ray Observatory and Rossi X-ray Timing Explorer.
These observations strongly suggest that
GRBs originate from cosmological distances. The fireball model 
(Goodman 1986; Paczy\'{n}ski 1986; Rees \& M\'{e}sz\'{a}ros 1992,1994;
M\'{e}sz\'{a}ros \& Rees 1992; M\'{e}sz\'{a}ros, Laguna \& Rees 1993; 
M\'{e}sz\'{a}ros, Rees \& Papathanassiou 1994; Katz 1994; Sari, Narayan
\& Piran 1996) becomes the most popular and successful model for GRBs,
although other models, such as a hypernova scenario proposed by
Paczy\'{n}ski (1997) can not be eliminated now.

After producing the main GRB, the cooling fireball is expected to
expand as a thin shell into the interstellar medium
(ISM) and generate a relativistic blastwave, although whether the 
expansion is highly radiative (Vietri 1997a,b) or adiabatic is
still controversial. Afterglows at longer 
wavelengths are produced by the shocked interstellar medium
(ISM) (M\'{e}sz\'{a}ros \& Rees 1997).
Much analytical work on GRB afterglows has been done (Waxman 1997a, b;
Tavani 1997; Sari 1997; M\'{e}sz\'{a}ros, Rees \& Wijers 1997;
Dai \& Lu 1997), and it has been found that for adiabatic expansion,
$R \propto t^{1/4}, \gamma \propto t^{-3/8}$, where $R$ is the
shock radius measured in the burster's static frame,
$\gamma$ is the Lorentz factor of the shocked ISM measured in the 
observer's frame and $t$ is the observed time. These scaling laws
are valid only at the ultrarelativistic expansion stage ($\gamma \gg 1$).

The purpose of this work is to study numerically the evolution of
adiabatic fireballs appropriate from the ultrarelativistic expansion stage
to the mildly relativistic expansion stage. We show that
although the scaling law between $\gamma$ and $t$ at the ultrarelativistic
expansion stage obtained in this work is the same as above,
our coefficient for $\gamma$ differs dramatically from that of
Waxman (1997a, b). A detailed set of equations are presented to calculate 
synchrotron radiation from the shocked ISM. We see that
radiation during the mildly relativistic phase ($2 < \gamma < 5$)
which was obviously neglected in the previous studies is quite important.

The structure of this paper is as follows. In Section 2 we 
calculate the dynamical evolution of the relativistic shock. The 
difference between our results and those of Waxman is
stressed. Synchrotron radiation from the shocked ISM is formulated 
and compared with observations in Section 3, and a brief discussion
is given in the final section.

\vspace{0.8cm}

\section{\bf DYNAMICAL EVOLUTION OF FIREBALLS}

A fireball with total initial energy $E_0$ and initial bulk Lorentz
factor $\eta \equiv E_0/M_0 c^2$, where $M_0$ is the
initial baryon mass and $c$ the velocity of light, is expected to 
radiate half of its energy in $\gamma$-rays during the GRB phase
(Sari \& Piran 1995), either due to an internal-shock or an 
external-shock mechanism (Paczy\'{n}ski \& Xu 1994; Vietri 1997b,
and references therein). Subsequently  the fireball will continue to 
expand as a thin shell into the ISM, generating an ultrarelativistic
shock.

The jump conditions for the shock can be described as
(Blandford \& McKee 1976):
\begin{equation}
    n'=(4 \gamma + 3) n,
\end{equation}
\begin{equation}
    e'=(4 \gamma + 3) \gamma  n m_p c^2,
\end{equation}
\begin{equation}
    \Gamma ^2 = \frac {(\gamma + 1)(4 \gamma - 1)^2}{8 \gamma + 10},
\end{equation}
where $n$ is the number density of the unshocked ISM, $n'$ and $e'$
are the number density and energy density of the shocked ISM in the
frame comoving with the shell, $m_p$ is the proton mass,
and $\Gamma$ is the Lorentz factor of the shock in the observer's frame.
These equations are valid for describing ultrarelativistic shocks
as well as mildly relativistic shocks (Blandford \& McKee 1976).
For $\gamma \gg 1$, we have $e' = 4 \gamma^2 n M_p c^2$ and
$\Gamma = \sqrt{2} \gamma$. We will assume that the shocked ISM
in the shell is homogeneous.

As usual, the expansion of the fireball
is thought to be adiabatic, during which energy is conserved, so we have
\begin{equation}
    4 \pi R^2 R \left(1 - {{v_s} \over {c}}\right) \gamma^2 e' 
= {{E_0} \over {2}} + f M_0 c^2,
\end{equation}
where $v_s$ is the observed velocity of the shock 
and $R (1 - v_s / c) \gamma$ is the thickness of the shocked ISM
in the comoving frame. $f$ is defined as
$f \equiv 4 \pi R^3 n m_p/(3 M_0)$. Using equation (1), 
equation (4) can be further expressed as:
\begin{equation}
3 \gamma^3 (1 - \sqrt {1-1/ \Gamma^2})(4 \gamma + 3) f = {\eta \over 2} + f.
\end{equation}
For $\gamma \gg 1$, equation (5) generates an approximate relation: 
$3 \gamma^2 f \approx \eta/2$, or equally,
\begin{equation}
\gamma^2 R^3 \approx M_0 \eta / (8 \pi n m_p) \equiv
E_0 / (8 \pi n m_p c^2).
\end{equation}

In order to study the evolution of $\gamma$, we should add two 
equations. Photons observed at a time interval of $\Delta t$ 
are in fact emitted at an interval of $\Delta t_b$:
\begin{equation}
   \Delta t_b = \frac {\Delta t}{1 - v/c} = \frac {\gamma \Delta t}{
       \gamma - \sqrt {\gamma^2 - 1}},
\end{equation}
where $t_b$ is
the time measured in the burster's static frame and  $v$ is the velocity
of the shocked ISM in the observer's frame, while $\Delta t_b$ and
variation of radius ($\Delta R$) are related due to 
\begin{equation}
    \Delta R = {\sqrt {\Gamma^2 - 1} \over \Gamma} c \Delta t_b.
\end{equation}

Equations (2), (3), (5), (7) and (8) present a perfect description of
propagation of the shock. Given initial values of $E_0$ and $M_0$,
$R(t)$ and $\gamma (t)$ can be evaluated numerically. We take 
$E_0 = 10^{52}$\,ergs, $n=1\,{\rm cm}^{-3}$, and $M_0 = 10^{-5}M_\odot$
and $2 \times 10^{-5}M_\odot$. Using
these equations, we numerically study the evolution of $\gamma$
and $R$ as functions of observed time. Our results are plotted
in Figs. 1 and 2, where only the dot-dashed line corresponds to
$M_0=10^{-5}M_\odot$, and the other lines to $M_0=2\times 10^{-5}M_\odot$.
It can be seen that $M_0$ only influences the early-time evolution
of the fireball.

Under the assumption that $\gamma \gg 1$,
we can derive a simple analytic solution,
\begin{equation}
    R^4 \approx R_0^4 + 4 k t,
\end{equation}
\begin{equation}
    \gamma \approx ({k \over 2cR^3})^{1/2},
\end{equation}
where $k=E_0/(4 \pi n m_p c)$, and $R_0$ is the initial value of $R$.
In Figs. 1 and 2, we also plot the
results of equations (9) and (10). It is clear that at early time, when 
$\gamma \gg 1$, the analytic solution is a quite good approximation
for the numerical results.

If $R \gg R_0$, that is $t \gg \tau$, where $\tau$ refers to the 
duration of the main GRB, then we can rewrite 
equations (9) and (10) as:
\begin{equation}
    R \approx 8.93 \times 10^{15}E_{51}^{1/4}n_1^{-1/4}t^{1/4} cm,
\end{equation}
\begin{equation}
    \gamma \approx 193 E_{51}^{1/8}n_1^{-1/8} t^{-3/8},
\end{equation}
where $E_0 =10^{51}E_{51}$ ergs, $n=1 n_1$ cm$^{-3}$ and $t$ is in units of
second. We find that
when $\gamma \gg 1$ and $t \gg \tau$, equations (11) and (12) also fit 
numerical results very well. Equations (11) and (12) are scaling laws
for $R$ and $\gamma$ mentioned in Introduction.

We next compare equation (12) with the previous studies.
The scaling law for $\gamma$ is quite simple, but the coefficient
should be treated with great carefulness, since it may affect 
synchrotron radiation and observational behaviors severely
(Sari 1997). Waxman (1997a,b) derived a result with a larger
coefficient: $\gamma \approx 332 E_{51}^{1/8}n_1^{-1/8} t^{-3/8}$. 
As shown in the next section, this will result in very strong 
radiation in both X-ray and optical bands.
We have plotted his results for $R (t)$ and $\gamma (t)$ in Figs. 1
and 2. The discrepancy is noticeable. The difference between his and
ours is due to the fact that he has used a relation
among $t$, $R$ and $\gamma$: $t = R / (2 \gamma^2 C)$,
which may be incorrect for the evolution of
ultra-relativistic adiabatic fireballs (Sari 1997), and we have adopted
another relation: $t = R / (8 \gamma^2 c)$, which is obtained by
integrating equations (7) and (8).

We want to emphasize that the equations in this section are
correct only for ultrarelativistic blastwaves ($\gamma \gg 1$) and mildly 
relativistic blastwaves ($2 < \gamma < 5$). This means that after about
three days our numerical results for the parameters used in the above might
be spurious. Of course, our calculation can be considerably prolonged
by adjusting the parameter ($E_{51}/n_1$).

\vspace{0.8cm}

\section{\bf X-RAY AND OPTICAL AFTERGLOWS}
 
\subsection{\bf Synchrotron radiation}

Electrons in the shocked ISM are highly relativistic. Inverse Compton cooling
of the electrons may not contribute to emission in 
X-ray and optical bands we are interested in. We
will consider only synchrotron radiation below. The electron 
distribution in the shocked ISM is assumed to be a power-law function
of electron energy, as expected for shock acceleration,
\begin{equation}
\frac {dn_e'}{d \gamma_e} \propto \gamma_{e}^{-p}, 
   \hspace{1cm} (\gamma_{min} \leq \gamma_e \leq \gamma_{max}),
\end{equation}
where $\gamma_{min}$
and $\gamma_{max}$ are the minimum and maximum Lorentz factors,
and $p$ is the index varying between 2 and 3. We suppose that the
magnetic field energy density (in the comoving frame) is a 
fraction $\xi_B^2$ of the energy density,
$B'^2/8 \pi = \xi_B^2 e'$, and that the electrons carry a fraction
$\xi_e$ of the energy. $\gamma_{min}$ is determined by 
\begin{equation}
\gamma_{min} = \xi_e \gamma \left(\frac{m_p}{m_e}\right)\left(\frac
{p-2}{p-1}\right).
\end{equation}
We estimate $\gamma_{max}$ by equating, as usual, the electron
acceleration timescale with the synchrotron cooling timescale,
and find (M\'{e}sz\'{a}ros, Laguna \& Rees 1993; Vietri 1997a) 
\begin{equation}
\gamma_{max} \approx 10^8 B'^{-1/2}.
\end{equation}

The spectral property of synchrotron radiation is clear
(Rybicki \& Lightman 1979).  In the
comoving frame, the characteristic photon frequency is 
$\nu_m = eB' \gamma_{min}^2 / (2 \pi m_e c)$, where $e$ is the electron
charge. The spectral peaks at $\nu_{max} \approx 0.29 \nu_m$. 
For frequency $\nu \gg \nu_{max}$, the flux density scales as
$S_{\nu} \propto \nu^{- \alpha}$, where $\alpha = (p-1)/2$, and
for $\nu \ll \nu_{max}$, $S_{\nu} \propto \nu^{1/3}$. Below we formulate
synchrotron radiation. 
First, using equation (1), we further express equation (13) as
\begin{equation}
\frac {dn_e'}{d\gamma_e} = C' \gamma_{e}^{-p},
\end{equation}
where 
\begin{equation}
C' = (p-1) \gamma_{min}^{p-1} (4\gamma+3)n.
\end{equation}
Second, the synchrotron radiation power emitted per unit volume is
\begin{equation}
j(\nu) \equiv \frac {dP(\nu)}{d\nu} = \frac {\sqrt{3} e^3B'C'}
     {m_e c^2} \int \limits_{\gamma_{min}}^{\gamma_{max}} 
     \gamma_{e}^{-p} F({{\nu} \over {\nu_c}}) d \gamma_e,
\end{equation}
with
\begin{equation}
F(x) = x \int \limits_{x}^{+\infty} K_{5/3}(t) dt,
\end{equation}
and 
\begin{equation}
\nu_c = \frac {\gamma_{e}^2 e B'}{2 \pi m_e c},
\end{equation}
where $K_{5/3}(t)$ is the Bessel function. The specific intensity at 
frequency $\nu$ in the comoving frame is thus written as
\begin{equation}
I_{\nu,co} = \frac {1}{4\pi} j(\nu) R(1-\frac {v_s}{c}) \gamma. 
\end{equation}
The observed frequency $\nu_{\oplus}$ and specific intensity
$I_{\nu_{\oplus},\oplus}$ are related to $\nu$ and $I_{\nu,co}$ 
due to the following relativistic translations (Rybicki \& Lightman 1979):
\begin{equation}
\nu_{\oplus} = (1+ v/c) \gamma \nu,
\end{equation}
\begin{equation}
I_{\nu_{\oplus},\oplus} = (1 + v/c)^3 \gamma^3  I_{\nu,co}.
\end{equation}

In previous researches, it is customary in analytic derivations to
consider that at a given time $t$ the emitting surface is located
at $2 \gamma^2 c t$ and that the disk seen by the observer has a 
radius $\gamma c t$, as it would be in the absence of the
deceleration (i.e., an ellipsoid). Then the relativistic fireball
expanding isotropically will produce an observed flux 
$ S_{\nu_{\oplus},\oplus} = \pi (\gamma ct)^2 I_{\nu_{\oplus},\oplus}/ D^2$,
where $D$ is the distance from the source to the Earth (Rees 1966). 
However, the deceleration dynamics was recently investigated in 
more detail and it was found that due to the deceleration the 
emitting surfaces become distorted ellipsoids, and at sufficiently
late times, most of the light (either bolometric or in a given 
band) comes from a ring-like surface (Waxman 1997c; Panaitescu \& 
M\'{e}sz\'{a}ros 1997). For a given observed frequency band, as the
shocked fluid is decelerated, the peak frequency 
($\nu_{peak} = \gamma \nu_{max}$) crosses the observed band, and the
region radiating in that band shrinks from the full disk to a narrow
ring. According to Panaitescu \& M\'{e}sz\'{a}ros (1997), at energies 
far below or above $\nu_{peak}$, the ratios of the equivalent radii 
of the emitting surfaces to $\gamma ct$ are approximately constant in time.
For example, these ratios are 2.8 and 3.3 
respectively, by assuming a homogeneous ISM and an adiabatic expansion. We
use the following equation to evaluate the observed flux density,
\begin{equation}
S_{\nu_{\oplus},\oplus} = \frac {\pi (k \gamma ct)^2 I_{\nu_{\oplus},\oplus}}
    {D^2},
\end{equation}
where $k = 2.8$ for $\nu_\oplus < \nu_{peak}$ and $k = 3.3$ for
$\nu_\oplus > \nu_{peak}$. Since $\nu_{peak}$ enters X-ray and
optical bands very quickly, the ``visible'' zone acts as a narrow 
ring for most of time and Equation (24) should be accurate 
enough for our calculations.
%The relativistic fireball expanding isotropically with
%the Lorentz factor $\gamma$, will produce an observed flux (Rees 1966)
%\begin{equation}
%S_{\nu_{\oplus},\oplus} = \frac {\pi \gamma^2(ct)^2 I_{\nu_{\oplus},\oplus}}
%    {D^2},
%\end{equation}
%where  $D$ is the distance from the source to the Earth.
So we get the observed flux density
\begin{equation}
S_{\nu_{\oplus},\oplus} = \frac {1}{4} (1 + \frac {v}{c})^3
    (1 - \frac {v_s}{c}) \gamma^{6} 
    \frac {R (kct)^2 j(\nu(\nu_{\oplus}))}{D^2}.
\end{equation}
This equation shows that $S_{\nu_{\oplus},\oplus}$ strongly depends on 
$\gamma$, so that the coefficient in Equation (12) should be treated
carefully (Sari 1997). 
The observed flux by a detector is an integral of
$S_{\nu_{\oplus},\oplus}$:
\begin{equation}
F_{\oplus} = \int \limits_{\nu_{l}}^{\nu_{u}} S_{\nu_{\oplus},\oplus}
     d \nu_{\oplus},
\end{equation}
where $\nu_u$ and $\nu_l$ are the upper and lower frequency limits of
the detector.

\vspace{0.8cm}

\subsection{\bf Comparison with observations}

Following the numerical solution in Section 2, we continue to calculate
the afterglow in X-ray and optical bands, using equation
(26). Some of the parameters are taken as follows:
$p=2.5$, $\xi_{e} = 0.5$, and $\xi_{B} = 0.1$, which are required by
the spectral and temporal properties of GRBs (Sari et al. 1997; 
Wijers, Rees \& M\'{e}sz\'{a}ros 1997).
The distance $D$ is set to be 3 Gpc. In order to get an X-ray flux, we 
integrate equation (26) from 0.1 keV to 10 keV, since it is approximately
in the bands available for BeppoSAX and ASCA.
As for optical flux, we use Equation (25) to calculate the R band
flux density ($S_R$).
%To obtain an optical
%flux, we also integrate equation (26) and calculate the mean flux density 
%($S_{OPT}$) between $3.4 \times 10^{14}$ and $5.8 \times 10^{14}$ Hz,
%which covers the optical I, R and V bands.
Our numerical results for X-ray flux ($F_X$) and $S_R$ are
illustrated in Fig. 3 and Fig. 4.
%For comparison, we have also calculated
%Waxman's results and plotted them in dotted lines.

Also plotted in Figs. 3 and 4 are some observed data,
which would make it possible for us to see to what extent the 
model could agree with observations. The X-ray data are quoted from: 
(1) Wijers, Rees \& M\'{e}sz\'{a}ros (1997); (2) Frontera et al. (1997);
(3) Costa et al. (1997a); (4) Butler et al. (1997); (5) Feroci et al.
(1997); (6) Piro et al. (1997a); (7) Costa et al. (1997d); (8) Piro et al.
(1997b); (9) Marshall et al. (1997); (10) Murakami et al. (1997a);
(11) Remillard et al. (1997); (12) Marshall et al. (1997b);
(13) Murakami et al. (1997b); and (14) Greiner et al. (1997).
Please note that since different detectors work in different bands and here
we have converted their flux data into $0.1-10$ keV band linearly, errors
to a factor of two are thus possible.
In Fig. 4 the observed R band flux densities are quoted from:
%The optical data are quoted from:
(1) Wijers, Rees \& M\'{e}sz\'{a}ros (1997); (2) Galama et al. (1997); 
(3) Fruchter et al. (1997).
%Here only I, R, V magnitudes
%are converted to flux densities in this paper, since these three
%frequencies are close to each other and are included in our 
%$3.4 \times 10^{14} - 5.8 \times 10^{14}$ Hz range.

Although the observed GRBs are expected to reside at different
distances and their intrinsic parameters such as $E_0$, $n$, 
$p$, $\xi_e$ and $\xi_B$ may vary markedly, it can be seen from 
Fig. 3 that the observed X-ray data are really quite easy to be 
reproduced by our model. The optical afterglow from GRB 970228
can be fitted quite well for $t \leq 3$ days. However after about three
days the theoretical light curve shows too sharp a
decline. As mentioned in Section 2, this might be spurious,
since our calculations are reliable only before about three days.
Because the X-ray afterglow of GRB970228 was observed $11.6$ days later
and the optical afterglow was observed more than $6$ months later, we suggest
that a marginally relativistic ($1.0005 < \gamma < 2$, or 
$10^4$ km/s $< v <2.6 \times 10^5$ km/s) model should be 
considered so as to provide a perfect description for the
afterglow. We noticed that Wijers et al.(1997) have pointed 
out that when the GRB remnant becomes nonrelativistic and enters
the Sedov-Taylor phase, the optical flux will decline as 
$F_{OP} \propto t^{(3-15 \alpha)/5}$, decaying faster than 
earlier time.

On 4 September the Hubble Space Telescope observed the afterglow 
of GRB 970228 for the third time and found that the optical 
transient has faded to $V \approx 28^m.0$ (Fruchter et al. 1997),
which corresponds to the R-band magnitude $\approx 27.0$, 
being well consistent with the power-law extrapolation of earlier data,
This suggests that the radiation during the whole period may be emitted
through one mechanism.
Although this seems to have confirmed the fireball model, we argue
below that it may be a puzzle. If supposing $\gamma$ dacays as 
$\gamma(t) = (200 \sim 300) (E_{51}/n_1)^{1/8} t^{-3/8}$, 
the relativistic condition $\gamma > 2$ for $t = 6.25$ months 
will require $(E_{51}/n_1) > (10^4 \sim 10^5)$, which seems quite
unlikely. This difficulty may be overcome by assuming that the 
ISM is not homogeneous, so that the shock can keep to be relativistic 
for more than $6$ months. Another possibility is that
$\gamma$ does fall below $2$ several days after the main GRB, but the
radiation during the marginally relativistic phase could account for
the long-term optical afterglow. This point should be investigated in 
more details.

The optical afterglow of GRB 970508 shows a first-rising-then-decreasing 
behavior (Castro-Tirado et al. 1997; Vietri 1997b; Djorgovski et al. 1997). 
The light curve peaks at about $t = 2$ days. This behavior can not be 
explained by the simple shock model described here, which only predicts 
a peak at $t = 10^3 \sim 10^4$ s. We suggest that an inhomogeneous ISM 
with some clumps might account for it. 
\vspace{0.8cm}

\section{\bf DISCUSSION}
BeppoSAX has made an important breakthrough in GRB researches. GRBs
are widely believed to be produced by relativistically expanding 
blastwaves, or namely fireballs, at cosmological distances.
Catastrophic events such as merging of
neutron star binaries (Narayan et al. 1992; Vietri 1996), failed supernovae
(Woosley 1993), collapse of magnetized white dwarfs (Usov 1992),
accretion-induced phase transitions of accreting neutron stars
(Cheng \& Dai 1996), and hypernovae (Paczy\'{n}ski 1997) have been suggested
as possible cosmological models of GRBs. One expects that extensive
observational and theoretical investigations on GRB afterglows should be
helpful to providing much more important clues to studies of the GRB origin.

We have shown in this paper that a relativistic fireball expanding
adiabatically into the uniform interstellar medium can roughly explain the 
afterglows of five observed GRBs, especially their X-ray flux. We have
made a detailed numerical study of the expansion and derived an approximate 
analytic solution. We would like to stress that the relation among $t$, $R$, 
and $\gamma$ should be $t = R / (8 \gamma^2 c)$ or 
$\Delta t = \Delta R / (2 \gamma^2 c)$. Our results indicate a smaller
coefficient for $\gamma (t)$:
$\gamma (t) \approx 193 (E_{51}/n_{1})^{1/8} t^{-3/8}$, which differs
noticeably from that of Waxman's (1997a,b). 
The difference in the coefficient results in great differences in 
observational effects. 
We present a set of equations
to describe the synchrotron radiation from the shocked ISM, and calculate
the X-ray and optical flux.
It is found that an expanding
fireball will no longer be highly relativistic ($\gamma \le 2$) about 3 
to 4 days after the main GRB. This leads to our suggestion that
a marginally relativistic expansion model analogous to (but still much 
faster than) that
for a supernova (Woosley \& Weaver 1986) should be established.

\vspace{5mm}
We would like to thank Dr. Ralph Wijers for many good suggestions 
and helpful comments that have led to an 
overall improvement of this paper.
This work was supported by the National Natural Science
Foundation, the National Climbing Programme on Fundamental
Researches and the Foundation of the Committee
for Education of China

\vspace{8mm}

\newpage
\baselineskip=4.5mm

\noindent
{\bf REFERENCES}

\noindent
\begin{description}
\item Blandford, R.D., and McKee, C.F. 1976, Phys. of Fluids, 19, 1130
\item Bond, H. et al. 1997, IAU Circ. 6654
\item Butler, R.C. et al. 1997, IAU Circ. 6539
\item Castro-Tirado, A.J. et al. 1997, IAU Circ. 6657
\item Cheng, K.S., and Dai, Z.G. 1996, Phys. Rev. Lett., 77, 1210
\item Costa, E. et al. 1997a, IAU Circ. 6533
\item Costa, E. et al. 1997b, IAU Circ. 6572
\item Costa, E. et al. 1997c, IAU Circ. 6576
\item Costa, E. et al. 1997d, IAU Circ. 6649
\item Costa, E. et al. 1997e, preprint: astro-ph/9706065
\item Dai, Z.G., and Lu, T. 1997, ApJ, submitted
\item Djorgovski, S.G. et al. 1997, Nat, 387, 876
\item Feroci, M. et al. 1997, IAU Circ. 6610
\item Fishman, G.J., and Meegan, C.A. 1995, ARA\&A, 33, 415
\item Frail, D.A. et al. 1997, IAU Circ. 6662
\item Frontera, F. et al. 1997, IAU Circ. 6637
\item Fruchter, A. et al. 1997, IAU Circ. 6747
\item Galama, T. et al. 1997, Nat, 387, 479
\item Goodman, J. 1986, ApJ, 308, L47
\item Greiner J. et al. 1997, IAU Circ. 6757
\item Groot, P.J. et al. 1997, IAU Circ. 6584
\item Heise, J. et al. 1997, IAU Circ. 6610
\item Katz, J. 1994, ApJ, 422, 248
\item Klebesadel, R.W., Strong, I.B., and Olson, R.A. 1973, ApJ, 182, L85
\item Marshall, F.E. et al. 1997a, IAU Circ. 6683
\item Marshall, F.E. et al. 1997b, IAU Circ. 6727
\item Metzger, M.R. et al. 1997, IAU Circ. 6655
\item Murakami, T. et al. 1997a, IAU Circ. 6687
\item Murakami, T. et al. 1997b, IAU Circ. 6632
\item M\'{e}sz\'{a}ros, P., Laguna, P., and Rees, M.J. 1993, ApJ, 415, 181
\item M\'{e}sz\'{a}ros, P., and Rees, M.J. 1992, MNRAS, 257, 29P
\item M\'{e}sz\'{a}ros, P., Rees, M.J., and Papathanassiou, H. 1994, ApJ,
      432, 181
\item M\'{e}sz\'{a}ros, P., and Rees, M.J. 1997, ApJ, 476, 232
\item M\'{e}sz\'{a}ros, P., Rees, M.J., and Wijers, R. 1997, ApJ, 
      submitted, preprint: astro-ph/9709273
\item Narayan, R., Paczy\'nski, B., and Piran, R. 1992, ApJ, 395, L83
\item Paczy\'{n}ski, B. 1986, ApJ, 308, L43
\item Paczy\'{n}ski, B. 1997, preprint: astro-ph/9706232
\item Paczy\'{n}ski, B., and Xu, G. 1994, ApJ, 427, 708
\item Panaitescu, A., and M\'{e}sz\'{a}ros, P. 1997, ApJL, 
      submitted, preprint: astro-ph/9709284
\item Piro, L. et al. 1995, Proc. SPIE, 2517, 169
\item Piro, L. et al. 1997a, IAU Circ. 6617
\item Piro, L. et al. 1997b, IAU Circ. 6656
\item Rees, M.J. 1966, Nat, 211, 468
\item Rees, M.J., and M\'{e}sz\'{a}ros, P. 1992, MNRAS, 258, 41P
\item Rees, M.J., and M\'{e}sz\'{a}ros, P. 1994, ApJ, 430, L93
\item Remillard, R. et al. 1997, IAU Circ. 6726
\item Rybicki, G.B., and Lightman, A.P. 1979, Radiative Processes in 
      Astrophysics, Wiley, New York
\item Sahu, K.C. et al. 1997, Nat, 387, 476
\item Sari, R. 1997, preprint: astro-ph/9706078
\item Sari, R., Narayan, R., and Piran, T. 1996, ApJ, 473, 204
\item Sari, R., and Piran, T. 1995, ApJ, 455, L143
\item Tavani, M. 1997, ApJ, 483, L87
\item van Paradijs, J. et al. 1997, Nat, 386, 686
\item Usov, V.V. 1992, Nat., 357, 472
\item Vietri, M. 1996, ApJ, 471, L91
\item Vietri, M. 1997a, ApJ, 478, L9
\item Vietri, M. 1997b, ApJ, 488, L105
\item Waxman, E. 1997a, ApJ, 485, L5
\item Waxman, E. 1997b, ApJL, in press, preprint: astro-ph/9705229
\item Waxman, E. 1997c, ApJL, submitted, preprint: astro-ph/9709190
\item Wijers, R., Rees, M.J., and M\'{e}sz\'{a}ros, P.
      1997, MNRAS, 288, L51
\item Woosley, S. 1993, ApJ, 405, 273
\item Woosley, S.E., and Weaver, T.A. 1986, ARA\&A, 24, 205
\end{description}

\newpage
\baselineskip=8mm

\noindent
{\large \bf Figure Captions}

\vspace{10mm}

\noindent
{\bf Figure 1.} Evolution of the shock radius. The solid line is our
numerical result and the dashed line our approximate analytic
solution (equation [9]). Waxman's result is plotted by the dotted line.
Only the dot-dashed line corresponds to
$M_0=10^{-5}M_\odot$, and the other lines to $M_0=2\times 10^{-5}M_\odot$.

\vspace{0.5cm}
\noindent
{\bf Figure 2.} Evolution of $\gamma$. The solid line is our
numerical result and the dashed line our approximate analytic
solution (equation [10]). Waxman's result is plotted by the dotted line.
Only the dot-dashed line corresponds to
$M_0=10^{-5}M_\odot$, and the other lines to $M_0=2\times 10^{-5}M_\odot$.

\vspace{0.5cm}
\noindent
{\bf Figure 3.} 0.1$-$10 keV flux ($F_X$) vs. time. $F_X$ is in
units of ergs cm$^{-2}$ s$^{-1}$. The full line is our 
numerical result.  For the observed fluxes, please see the text.

\vspace{0.5cm}
\noindent
{\bf Figure 4.} Optical flux density ($S_R$) vs. time.
$S_{R}$ is in units of ergs cm$^{-2}$ s$^{-1}$ Hz$^{-1}$. The
solid line is our numerical result.  For the observed flux densities,
please see the text.

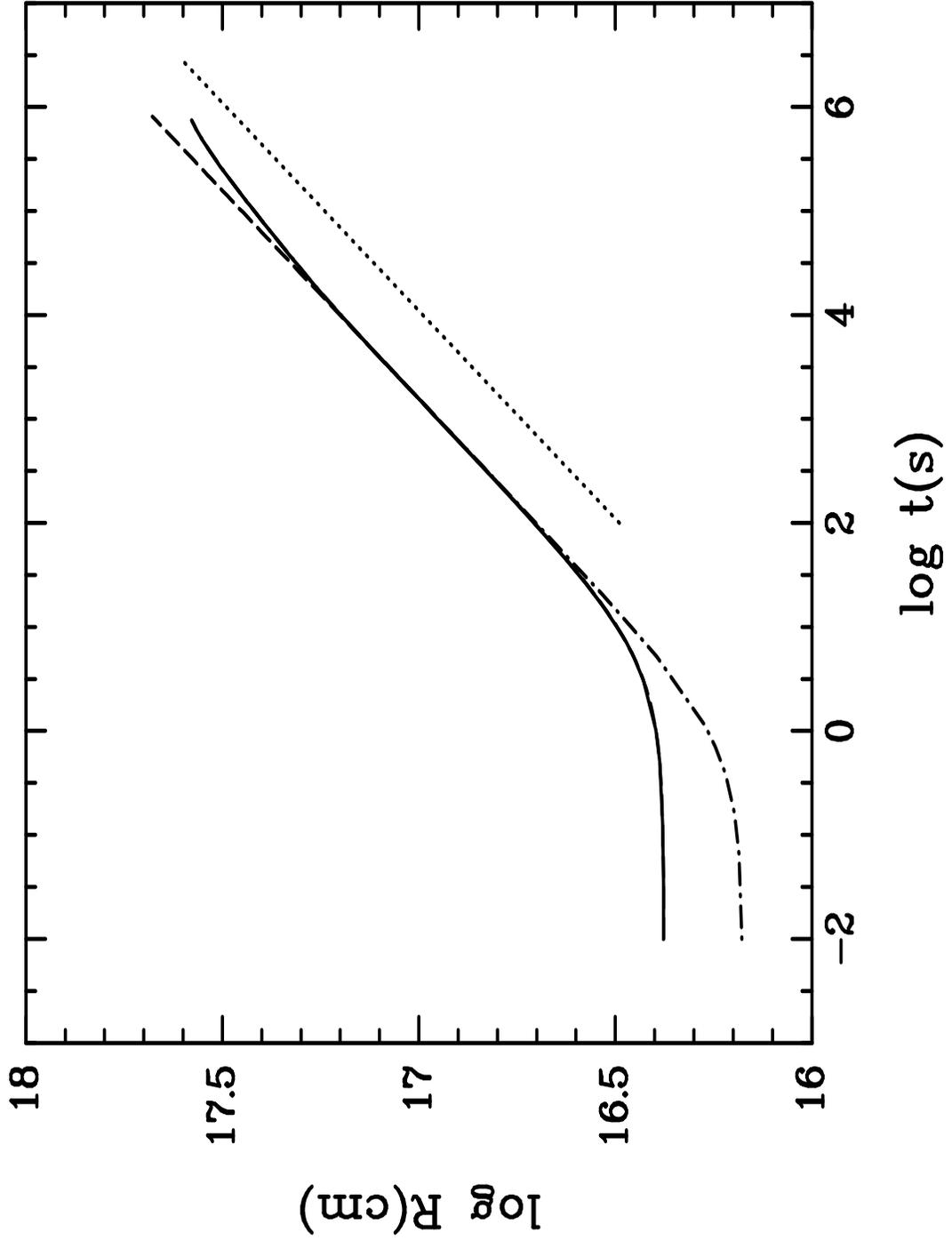
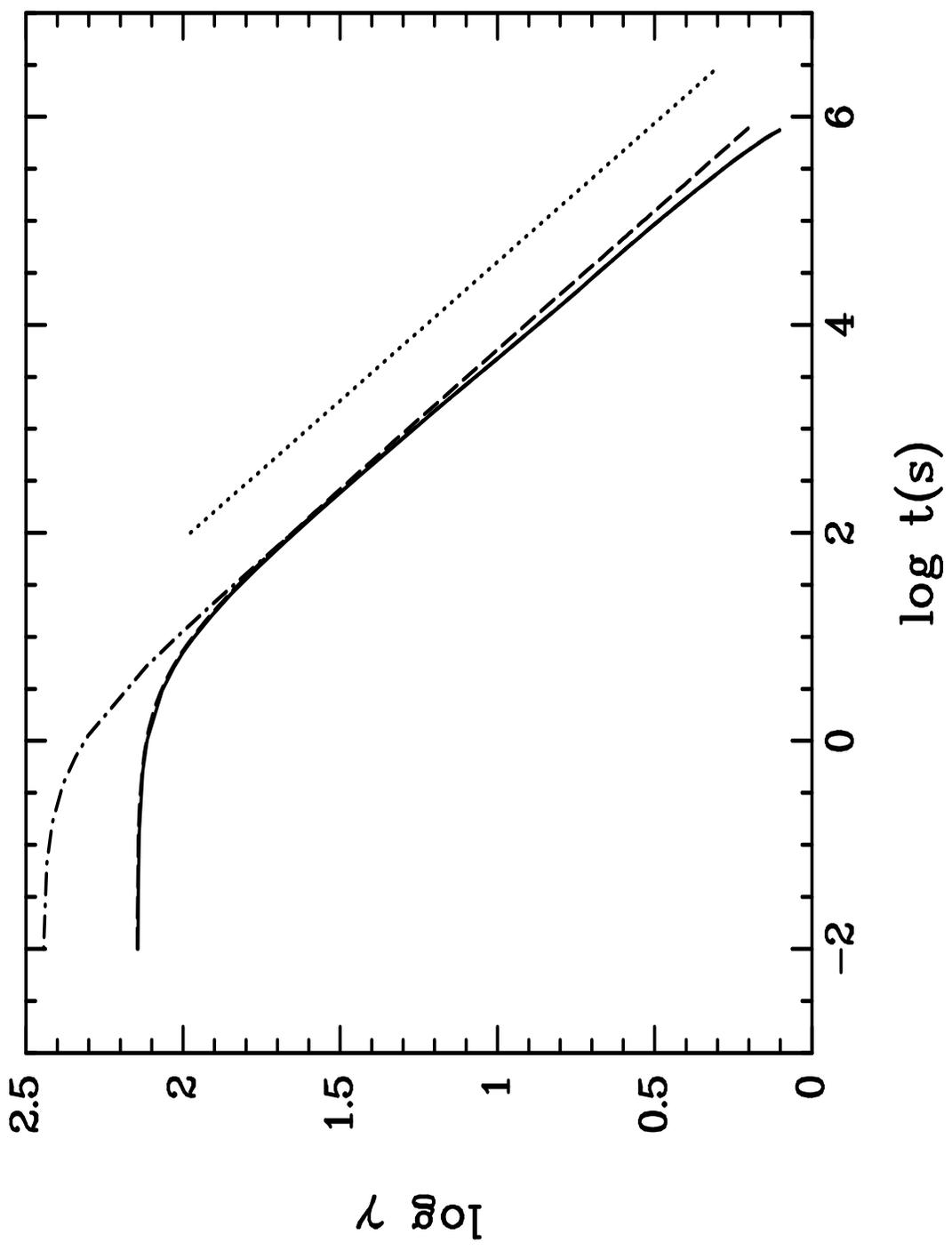
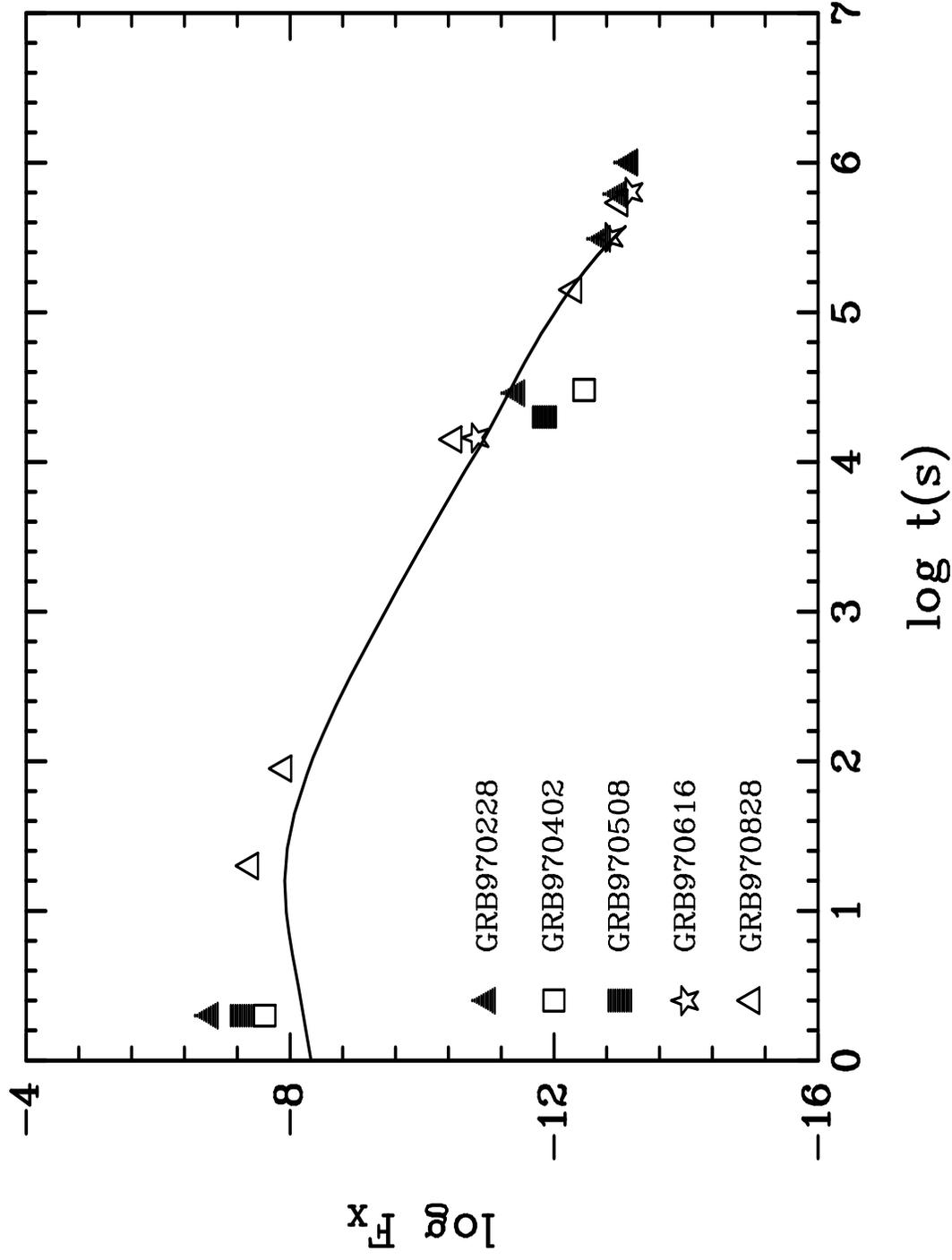
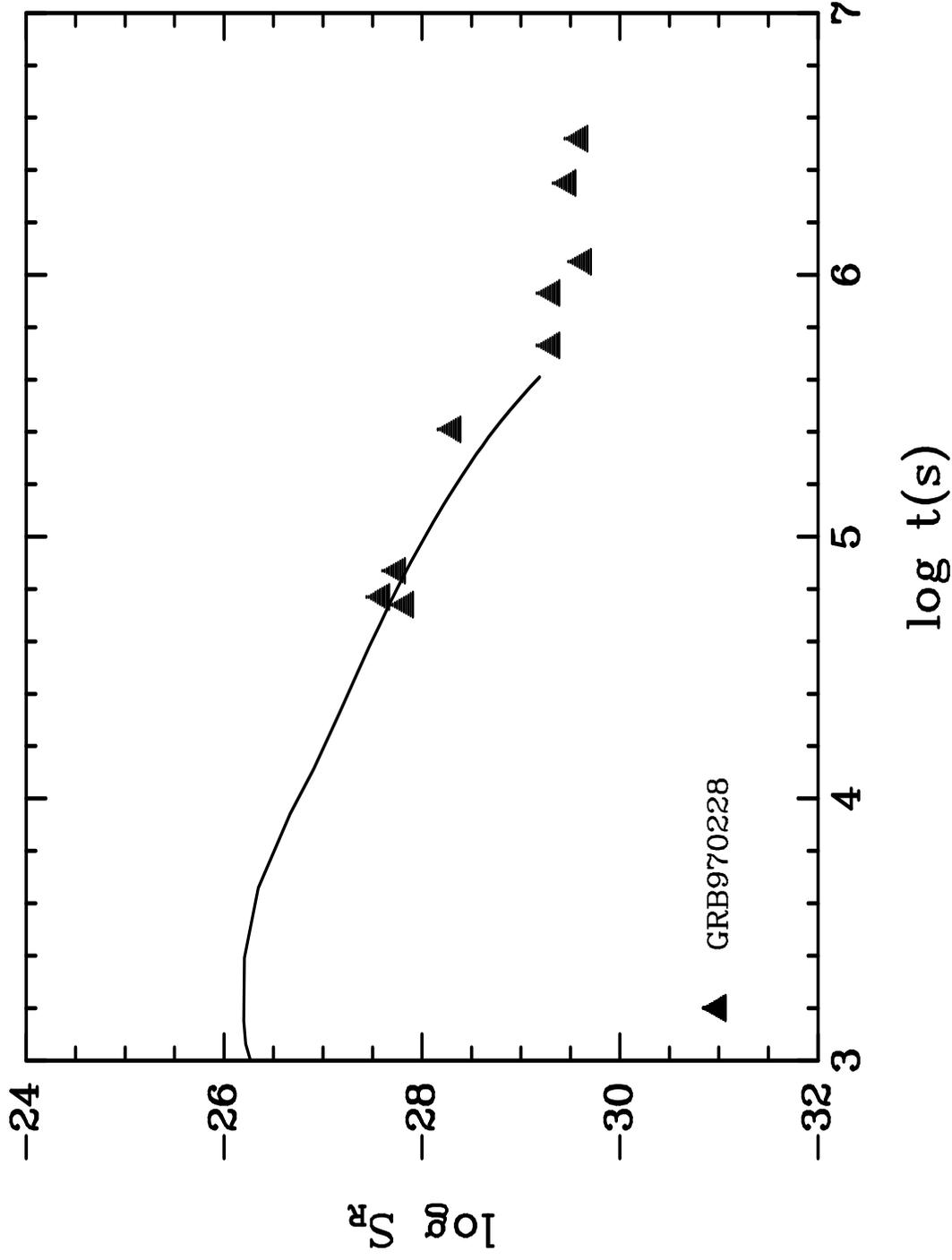
\end{document}